 \documentclass{aa}
\usepackage{graphicx}
\usepackage{natbib}
\bibpunct{(}{)}{;}{a}{}{,}

\usepackage{txfonts}
\begin{document}

\title{Identification of different types of kink modes in coronal loops: principles and application to TRACE results}

\author{T. J. Wang\inst{1,2},  S. K. Solanki\inst{3} \and M. Selwa\inst{1}}

     \offprints{T.J. Wang}

     \institute{ Department of Physics, Catholic University of America and 
NASA Goddard Space Flight Center, Code 671, Greenbelt, MD 20771\\                 
                  \email{wangtj@helio.gsfc.nasa.gov}
          \and
              Department of Physics, Montana State University, 
                  Bozeman, MT 59717-3840, USA\\ 
          \and
                  Max-Planck-Institut f\"{u}r Sonnensystemforschung,
                  37191 Katlenburg-Lindau, Germany\\
             }

\date{Received ----; accepted ----}

\abstract
{We explore the possible observational signatures of different types of kink modes (horizontal and vertical 
oscillations in their fundamental mode and second harmonic) that may arise in coronal loops, with the aim 
of determining how well the individual modes can be uniquely identified from time series of images. 
A simple, purely geometrical model is constructed to describe the different types of kink-mode oscillations.
These are then `observed' from a given direction. In particular, we employ the 3D geometrical parameters 
of 14 TRACE loops of transverse oscillations to try to identify the correct observed wave mode.
We find that for many combinations of viewing and loop geometry it is not straightforward to distinguish 
between at least two types of kink modes just using time series of images. We also considered Doppler 
signatures and find that these can help obtain unique identifications of the oscillation 
modes when employed in combination with imaging. We then compare the modeled spatial signatures with the 
observations of 14 TRACE loops.  We find that out of three oscillations previously identified as fundamental 
horizontal mode oscillations, two cases appear to be fundamental vertical mode oscillations (but possibly
combined with the fundamental horizontal mode), and one case
appears to be a combination of the fundamental vertical and horizontal modes, while in three cases it is not 
possible to clearly distinguish between the fundamental mode and the second-harmonic of the horizontal 
oscillation. In five other cases it is not possible to clearly distinguish between a fundamental 
horizontal mode and the second-harmonic of a vertical mode.

}

\titlerunning{Identification of different types of kink modes in coronal loops}
\authorrunning{T.J. Wang et al.}
\maketitle

\section{Introduction}
The solar corona is characterized by highly dynamic loop-like structures believed to outline magnetic 
field lines. Various types of coronal loop oscillations have been observed for decades in radio, visible, 
EUV, and X-rays \citep[see, e.g., reviews by][]{nak03, nak05, asc04, wan04a, wan05a}. Recently, 
temporally and spatially resolved transverse and longitudinal oscillations have been detected in 
coronal loops in the high-resolution observations by the Transition Region And Coronal Explorer (TRACE) 
and the Solar and Heliospheric Observatory (SOHO). Compressible longitudinal waves in cool ($\sim1$ MK) 
loops were first detected by the EUV Imaging Telescope (EIT) on board SOHO \citep{ber99}, which were 
confirmed by TRACE observations \citep[e.g.][]{dem00,dem02a,dem02b}, and identified as propagating 
slow magnetoacoustic waves \citep{nak00}. Global kink-mode oscillations were first found in cool 
($\sim$ 1~MK) coronal loops from TRACE EUV imaging observations \citep{asc99,asc02,nak99}. They were 
seen as spatial displacements with periods of 3$-$5 minutes and were apparently excited by flares or 
erupting filaments \citep{asc02,sch02}. Later, strongly damped standing slow-mode waves were discovered 
with the Solar Ultraviolet Measurement of Emitted Radiation (SUMER) spectrometer on-board SOHO by virtue 
of their Doppler signatures in the flare lines Fe~{\small XIX} and Fe~{\small XXI} 
\citep[e.g.][]{wan02, wan03a, wan03b, ofm02}. These oscillations are usually set up immediately following 
a pulse of hot plasma flowing from one of the loop footpoints, associated with small flares or even micro-flares 
\citep{wan05b, wan06}. Extensive studies on coronal loop oscillations in recent years have made 
{\it MHD coronal seismology} a practical, new tool for the determination of previously poorly known 
parameters of the coronal structure \citep[see, e.g.,][for reviews]{rob84, rob03, nak05}. For example, 
the measured oscillation period has been applied to determine the mean magnetic field strength of coronal 
loops in case of either kink mode \citep{nak01, ver04} or slow mode oscillations \citep{wan07}. 
\begin{figure*}
\centering
\includegraphics[width=16cm]{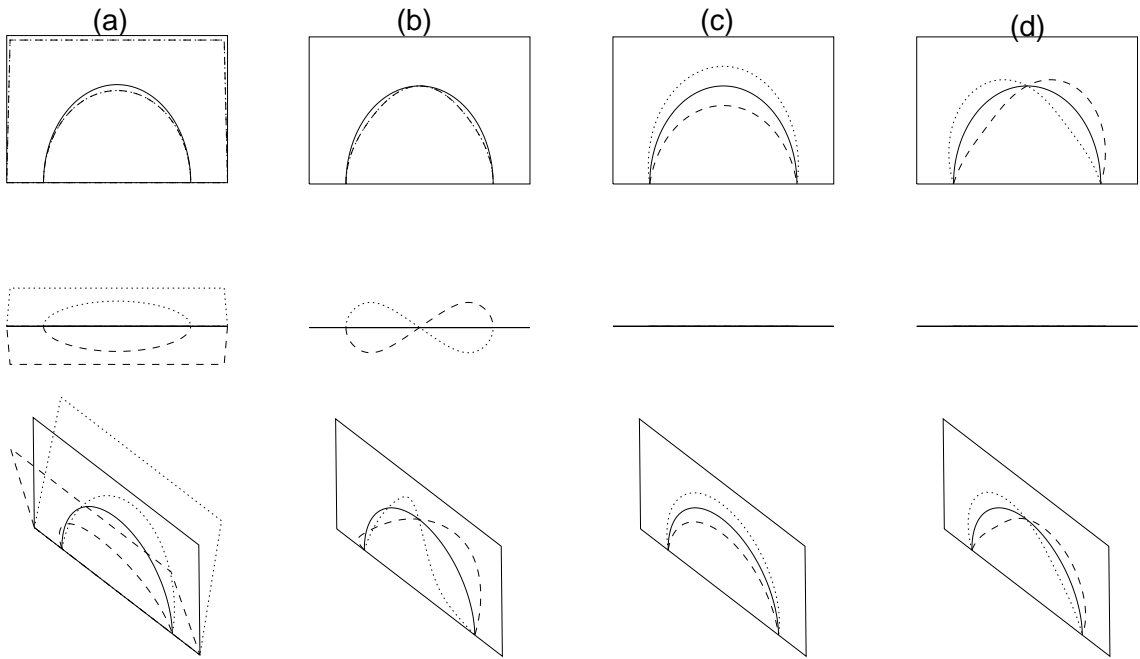}
\caption{ A sketch of the horizontal and vertical oscillations and their second harmonics. 
(a) The fundamental horizontal mode, (b) the second harmonic horizontal mode, (c) the fundamental 
vertical mode, and (d) the second harmonic vertical mode. The top, middle and bottom rows correspond
to views from the side, top and top-left, respectively.  }
\label{sketch}
\end{figure*}

\citet{wan04b} first suggested that the longitudinal curvature of loops may lead to
two types of kink modes (i.e., horizontal and vertical modes) in a coronal loop based on observations. 
The motion is termed a {\it horizontal} mode
if the loop rocks perpendicularly to the loop plane. If the oscillation is polarized in the loop plane
and manifests expanding and shrinking motions, we refer to it as a {\it vertical} mode.
The horizontal kink mode oscillation for an individual loop has been numerically
studied by \citet{miy04} with a 3D MHD model. They set an initial horizontal velocity field  
near the loop top as a perturbation to induce the oscillation. The vertical mode oscillation can be excited 
either by a pressure pulse below the loop apex\citep[e.g.][] {sel05, sel06, sel07} or by a velocity driver
in one of the footpoints \citep[e.g.][] {bra05, bra06, ver06a, ver06b} as simulated in a 2D 
arcade model. Very recently, \citet{mcl08} performed a 3D nonlinear MHD simulation of wave activity 
in active regions in which individual loop density structure is included. They found that the impact 
of the fast wave impulsively excites both horizontal and vertical loop oscillations. 
 \citet{van04} first studied the eigenmodes of a curved loop and showed that curvature only slightly
affects the damping times of the observed kink oscillations in comparison with a straight tube.
By numerically solving the eigenvalue problem of a similar model, \citet{ter06} found that 
curvature and density structuring introduce preferential directions of oscillation showing the 
existence of two kink eigenmodes with horizontal and vertical polarizations, but their frequencies
and damping rates are very similar. Their calculation supports the initial notion of \citet{wan04b}.

In several examples, the second harmonic was found together with the fundamental mode via
time series analysis \citep{ver04, van07}. The first spatially resolved example of the second 
harmonic of transverse loop oscillations was recently identified by \citet{dem07}. 
\citet{dia06} and \citet{dia06b} analytically studied the vertical kink mode and its second 
harmonic (they termed it the {\it swaying} kink mode). Observations of multiple periodicities in kink 
loop oscillations \citep[e.g.,][]{ver04} have been used to extract information
on the internal longitudinal structuring of coronal loops by \citet{and05, mce06} 
and has been demonstrated to have high potential value for coronal seismology. In addition, 
\citet{ter07} theoretically studied the excitation of the eigenmodes of coronal loops due to an initial
disturbance. Their results suggest that longitudinal harmonics of kink modes are in principle 
more easily excited than azimuthal harmonics (fluting modes).

\begin{figure*}
\centering
\includegraphics[width=12cm]{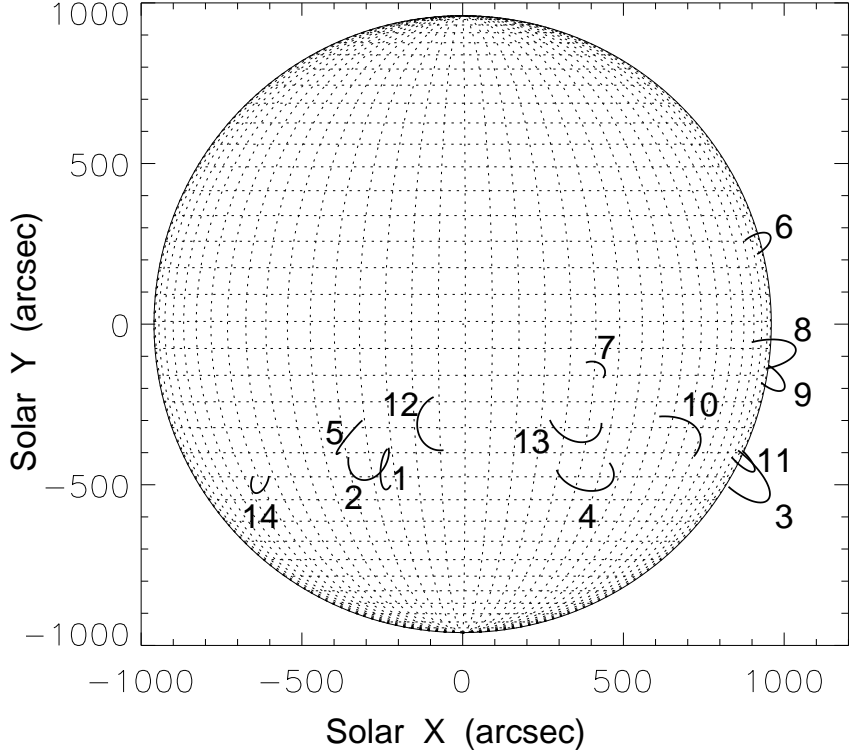}
\caption{Positions of the 14 studied oscillating TRACE loops on the solar disk.}
\label{loops}
\end{figure*}

Therefore, differentiating between various types of kink modes and their higher harmonics
in observations will improve 
the applications of coronal seismology in precisely determining the unknown physical 
parameters of the coronal plasma. \citet{asc02} identified oscillations in 26 coronal loops as  
kink modes. They did not distinguish between the different types of modes. \citet{wan04b} have
for the first time identified the vertical kink mode oscillation of a TRACE loop above the limb based on 
its spatial signature, which is distinct from the horizontal mode.  Our aim here is to explore 
observational signatures of different types of kink-mode oscillations based on the loop geometries measured by
\citet{asc02} and check if each observed oscillation can be identified with a given type of 
kink mode based on these signatures.

In Section 2, we describe the methods which are used to model the horizontal and vertical 
kink modes. In Section 3, we discuss their similarities and differences in signatures of the 
spatial pattern and Doppler shift. In Section 4, we discuss the simultaneous presence of multiple 
(harmonics) modes. Conclusions are summarized in Section 5.

\section{Modeling of different types of kink mode oscillations}
\label{smsct}
In principle we need to solve the three-dimensional MHD wave equations to first obtain
the eigenfunction and then use it to analyze the motion of the eigenmodes of coronal loops 
\citep[e.g.,][]{van04, ter06}. Since the aim of this paper is to provide a simple guide to interpret 
observations, we model  
the loop oscillations based on simple geometric considerations. The loops are assumed to be symmetric 
circular arches lying in a plane, but it is cautioned that in general real loops do not satisfy 
this condition. We set up a Cartesian coordinate
system ($x$, $y$, $z$) with the origin at the midpoint between the two footpoints of the loop. The $x$-axis
is along the loop baseline, and the $z$-axis is directed towards the apex of the loop. The position of 
a point along the loop is first transformed from the Cartesian system into the spherical polar system,
with the coordinates ($R$, $\phi$, $\psi$), where the $y$-axis is taken to be the axis of rotation. $R$, $\phi$
and $\psi$ are the radius, latitude and longitude, respectively.

We model horizontal oscillations by modulating the latitude $\phi$ in the form,
\begin{equation}
 \phi(\psi, t)=\phi_m{\rm cos}\psi{\rm sin}\omega_1{t} {\rm ~~~(for~fundamental~mode)}, 
 \label{eqhf}
\end{equation}
or
\begin{equation}
 \phi(\psi, t)=\phi_m{\rm sin}2\psi{\rm sin}\omega_2{t} {\rm ~~~(for~second~harmonic)},
 \label{eqhs}
\end{equation}
where $\psi$ runs between $-\pi/2$ and $\pi/2$ for a point along the loop between its two footpoints,
$\phi_m$ is a small constant, representing the maximum amplitude, $\omega_1$ and $\omega_2$ are the
oscillation frequencies for the fundamental mode and the second harmonic. The equilibrium position 
of the loop is given when $\phi_m=0$. Figures~\ref{sketch}a and~\ref{sketch}b illustrate the fundamental 
mode and second harmonic horizontal oscillations of a semi-circular loop, modeled by taking 
$\phi(\psi, t)$ as the form given by Eq.~(\ref{eqhf}) and Eq.~(\ref{eqhs}), respectively.

We model vertical oscillations by modulating the radius $R$,
\begin{equation}
 R(\psi, t)=R(\psi)(1+A_m{\rm cos}\psi{\rm sin}\omega_1{t}) {\rm ~~~(for~fundamental~mode)},
 \label{eqvf} 
\end{equation}
or
\begin{equation}
 R(\psi, t)=R(\psi)(1+A_m{\rm sin}2\psi{\rm sin}\omega_2{t}) {\rm ~~~(for~second~harmonic)},
 \label{eqvs} 
\end{equation}
where $A_m$ is a small constant representing the relative maximum displacement amplitude, 
$\omega_1$, $\omega_2$, and $\psi$ have the same definitions as in Eqs.~(\ref{eqhf}) and (\ref{eqhs}). 
Figures~\ref{sketch}c and~\ref{sketch}d illustrate 
the fundamental mode and second harmonic vertical oscillations of a semi-circular loop, modeled by 
taking $\phi(\psi, t)$ as the form given by Eq.~(\ref{eqvf}) and Eq.~(\ref{eqvs}), respectively.

\begin{figure*}
\centering
\includegraphics[width=16cm]{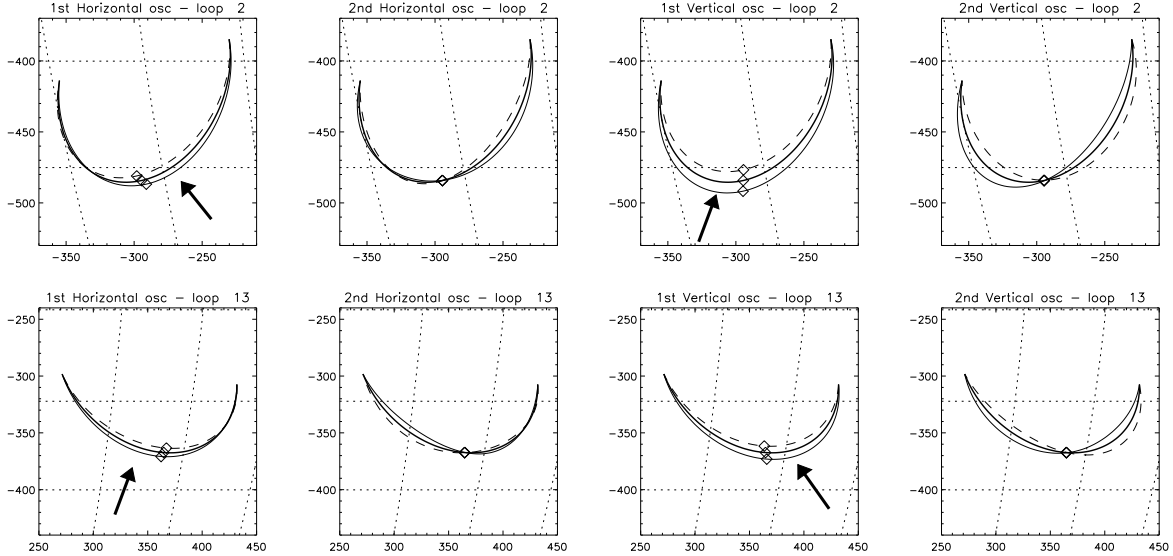}
\caption{Comparison between the fundamental mode ({\em left}) and the second harmonic ({\em middle left}) 
of the horizontal oscillation as well as the fundamental mode ({\em middle right}) and the second harmonic 
({\em right}) of the vertical oscillation for loops with the first category of geometries (loop 2: top row, 
loop 13: bottom row). The thick solid lines represent the equilibrium position of the oscillating
loop and the thin solid and dashed lines represent the two extrema at opposite phases of the oscillation. 
The apex position of the loop at each snapshot is indicated by a {\em diamond}. Arrows indicate the location
of maximum displacement due to the horizontal and vertical fundamental oscillation.}
\label{type1}
\end{figure*}

\citet{dia06} analytically studied oscillation modes of a
2D semi-circular magnetic arcade. They named the solution of the wave equation with an azimuthal 
wavenumber of $m=1$ the vertical mode and the solution with $m=2$ as the swaying mode. The
fundamental vertical and the swaying modes correspond to the fundamental and
second harmonic kink modes, respectively, in the straight flux slab or cylindrical tube
\citep{edw82, edw83}.

\section{The modeled signatures for 14 TRACE loops}
\label{trcsct}
\subsection{Observations}
Transverse oscillations of 26 coronal loops observed by TRACE in the 171 and 195 \AA~wavelength
bands have been analyzed by \citet{asc02}. They interpreted these 
oscillations in terms of fast kink modes, but did not consider their possible classification.
Here we select 14 of these loops (see Table~\ref{partab}) to model the horizontal and vertical 
oscillations in their fundamental mode and second harmonic in order to discuss if the types of these 
kink oscillations can be uniquely identified based on their signatures in TRACE movies. We chose 
these examples because they were illustrated by \citet{asc02} in their paper and we can easily 
compare our modeled results with their figures. As we shall show 
in Sect.~\ref{spcsct}, imaging alone often does not allow a unique identification. 
We therefore also consider if the spectroscopic signature of these modes is helpful 
to resolve remaining ambiguities (Sect.~\ref{dplsct}). We employ the geometrical parameters 
of the loops as measured by \citet{asc02}. Figure~\ref{loops} shows the orientations and positions 
of the sampled loops on the solar disk at the time of the observed oscillation. The loops' 
geometrical parameters are listed in Table~\ref{partab}.

\subsection{Spatial features}
\label{spcsct}
\begin{figure*}
\centering
\includegraphics[width=16cm]{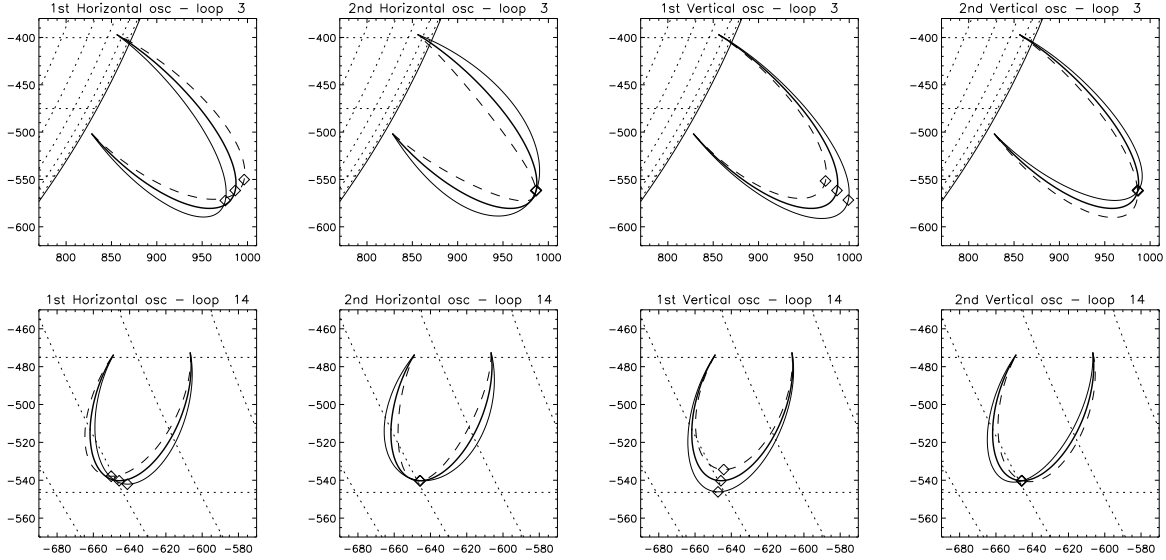}
\caption{Same as Fig.~\ref{type1} but for loops with geometries belonging to the second category  
(loop 3 on top and loop 14 at the bottom).}
\label{type2}
\end{figure*}

\begin{figure*}
\centering
\includegraphics[width=16cm]{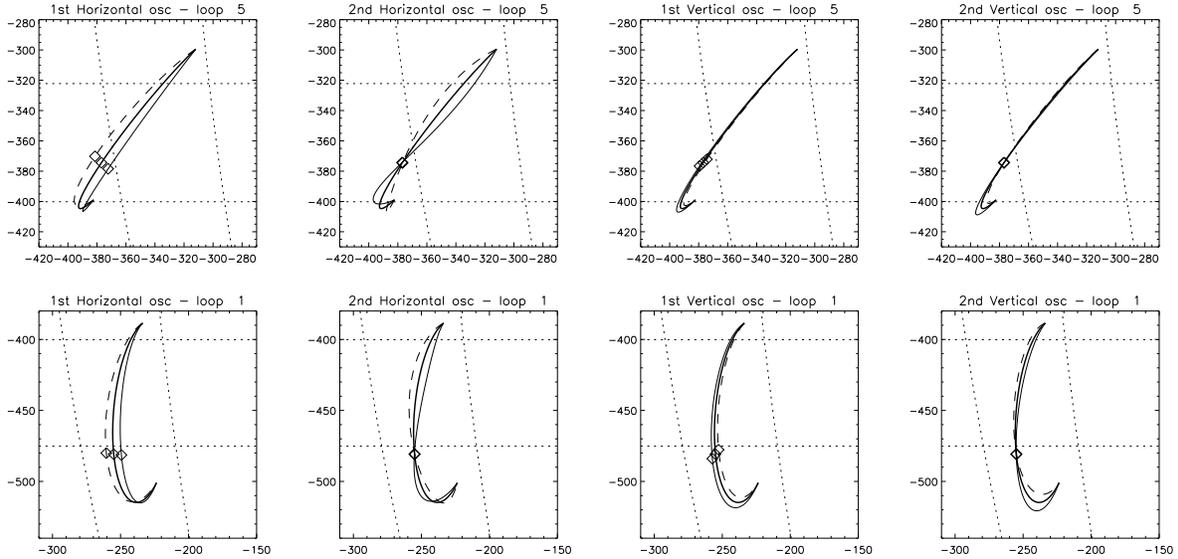}
\caption{ Same as Fig.~\ref{type1} but for loops with geometries belonging to the third category:
loop 5 in the top row and loop 1 in the bottom row.}
\label{type3}
\end{figure*}

 For the 14 loops, we model the horizontal oscillation with an amplitude of $\phi_m=5^{\circ}$ (or 
0.087 arc deg) and model the vertical oscillation with a relative amplitude of $A_m=0.087$ 
(see Sect.~\ref{smsct}). Note that we set the same maximum relative amplitudes for the horizontal
and vertical oscillations in order to make reasonable comparisons. 
Based on the similarity and difference in the projected signatures of the four types of kink modes, 
the 14 loops can be divided into three categories. Some typical cases are shown in 
Figs.~\ref{type1}$-$\ref{type3} and the main characteristics are summarized in Table~\ref{cmptab}.

\noindent
{\em -- Category I (Loops 2, 4, 7, 10, 12 and 13):}\\ 
Loops in this category are located close to disk center and their loop planes have a large inclination angle 
($\gamma$) to the line-of-sight direction. The values of $\gamma$ are in the range $37^{\circ}-63^{\circ}$
(Table~\ref{partab}). Figure~\ref{type1} shows two typical examples of such loops. For each of these the
observational signature of all four kink modes is illustrated. At first sight the horizontal
oscillations have a similar appearance to the vertical oscillations. In the case of the fundamental mode they 
both exhibit an evident displacement around the loop top in the projected plane (irrespective of whether
it is the horizontal or the vertical fundamental mode), while in the case of the
second harmonic they show a node at the loop top and maximum displacements at the legs. 
However, if we scrutinize the position of the point on the loop with the maximum projected displacement (indicated
by arrows in the first and third column panels of Fig.~\ref{type1}) for the fundamental mode, we find 
that this point is located on different sides of the loop apex for the horizontal 
and vertical oscillations  (e.g., to the right of the apex of loop 2 for the horizontal 
mode and to the left for the vertical mode; see arrows in the top first and third panels of Fig.~\ref{type1}). 
Therefore, we may identify the type of the observed kink oscillations according to
the measured position of the most evident displacements. The locations of the most pronounced oscillation 
for the studied cases have been marked in Figures 3$-$16 of \citet{asc02}. Thus, the observation of transverse 
oscillations of loop 2 is better explained by the fundamental horizontal mode. For loops 12 (not shown) 
and 13 (bottom panels of Fig.~\ref{type1}), the location of the most pronounced oscillation from \citet{asc02}
supports them as the fundamental vertical mode. Note that the case of loop 12 was re-examined by \citet{dem07}
recently, who identified the oscillation as the second harmonic horizontal mode based on the signature of 
an apparent node, whereas \citet{asc02} measured the oscillation for the top half of the loop by considering 
the apparent node to be a footpoint, and interpreted it as the fundamental kink mode. Since the measured 
oscillation period is peculiar long and better matches the fundamental mode, the interpretation by the 
second harmonic remains debated.  

For the second harmonic, the displacements are
evident in only one leg for the horizontal oscillation while they are evident in both legs for the vertical
oscillation. The observations of loops 4, 7 and 10 all show oscillations only in one leg,
so that they are consistent with the signature of the second harmonic horizontal mode. However, we notice 
that the fundamental horizontal oscillations also shows an evident displacement mainly in one loop leg, so that
we cannot clearly distinguish between these two modes, even when we take into account measurements of their 
periods. For example, \citet{asc02} measured an oscillation period of 136 s and a loop length of 113 Mm for 
loop 10, yielding an estimate of the phase speed of 830 km~s$^{-1}$ for the second harmonic which is in the range 
of acceptable values.

\begin{table*}
 \begin{center}
\caption{Geometrical parameters of the analyzed TRACE oscillating loops$^{\mathrm{a}}$, which are taken
from Aschwanden et al. (2002).
}
\label{partab}
\begin{tabular}{lccccccccccc}
\hline
  Loop & $l_0-l_c$ & $b_0-b_c$ & ${\alpha}$ & ${\theta_0}$ &  $h_0$ &  $r$ & $\gamma$ & Aschwanden & Geom. & Likely & Multiple \\
  No   &  (deg)    &   (deg)   &  (deg)     &  (deg) &  Mm  & (Mm) &  (deg) & et al. (2002)& Category & Mode$^{\mathrm{b}}$ & Modes\\ 
\hline
1  & $-$15.6 & $-$27.6 & 87  & 7     & 9    & 47  &20  & 1a, Fig.3   & III & 1H          & \\
2  & $-$19.6 & $-$24.5 & 12  & $-$44 & 11   & 57  &54  & 1f, Fig.4   & I   & 1H          & \\
3  & 82.3    & $-$27.7 & 152 & $-$12 & 38   & 99  &25  & 3a, Fig.5   & II  & 1H or 2V    & \\
4  & 26.0    & $-$27.3 & 7   & $-$14 & 12   & 74  &40  & 4a, Fig.6   & I   & 1H or 2H    & \\
5  & $-$22.9 & $-$21.3 & 47  & 2     & 0    & 53  &7   & 5c, Fig.7   & III & 1H          & \\
6  & 70.7    & 16.4    & 158 & 22    & 28   & 43  &30  & 7a, Fig.8   & II  & 1H,2H or 2V & \\
7  & 25.6    & $-$8.6  & 143 & 53    & $-$8 & 30  &54  & 8a, Fig.9   & I   & 1H or 2H    & \\
8  & 72.6    & $-$3.8  & 157 & 20    & 67   & 77  &25  & 10a, Fig.10 & II  & 1H or 2V    & \\
9  & 78.9    & $-$6.3  & 25  & $-$30 & 43   & 47  &26  & 11a, Fig.11 & II  &             & 1H+1V \\
10  & 49.2   & $-$20.2 & 150 & 49    &$-$71 & 120 &37  & 12b, Fig.12 & I   & 1H or 2H    & \\
11  & 75.2   & $-$21.8 & 10  & $-$22 & 42   & 48  &19  & 14a, Fig.13 & II  & 1H,2H or 2V & \\
12  & $-$4.7 & $-$18.8 & 99  & 63    &$-$10 & 65  &63  & 15a, Fig.14 & I   & 1V$^{\mathrm{c}}$ & 1H+1V \\
13  & 22.7   & $-$18.3 & 177 & $-$39 &$-$11 & 68  &49  & 16a, Fig.15 & I   & 1V                & 1H+1V \\
14 & $-$48.7 & $-$28.0 & 1   & $-$41 & 19   & 33  &38  & 17a, Fig.16 & II  & 1H or 2V    & \\
\hline
\end{tabular}
\end{center}
\begin{list}{}{}
\item[$^{\mathrm{a}}$]
$l_0-l_c$ and $b_0-b_c$ are the heliographic longitude and latitude relative to Sun center 
for the midpoint of the loop footpoint baseline. $\alpha$ is the azimuth angle of the loop baseline
to the east-west direction. $\theta_0$ is the inclination angle of the loop plane to
the vertical. $h_0$ is the height of the center of the circle describing the loop in the loop
plane. $r$ is the radius of the circular loop. $\gamma$ is the inclination angle of the loop 
plane to the line-of-sight direction. The column named ``Aschwanden et al. (2002)" lists the case and
corresponding figure numbers in that paper. The last third column lists the category of viewing
geometry for loops. The last two columns list the type of likely single and multiple kink modes identified
by comparing the observed signature with our geometrical model.
\item[$^{\mathrm{b}}$] 1H and 2H indicate the fundamental and the second-harmonic horizontal oscillations,
respectively, 1V and 2V the fundamental and the second-harmonic vertical oscillations. 1H+1V indicates 
a combination of the fundamental horizontal and vertical modes.
\item[$^{\mathrm{c}}$] This case was recently re-examined and identified as the second harmonic 
horizontal mode by \citet{dem07}, based on the presence of an apparent node (see the text). 
\end{list}
\end{table*}

\begin{figure*}
\centering
\includegraphics[width=16cm]{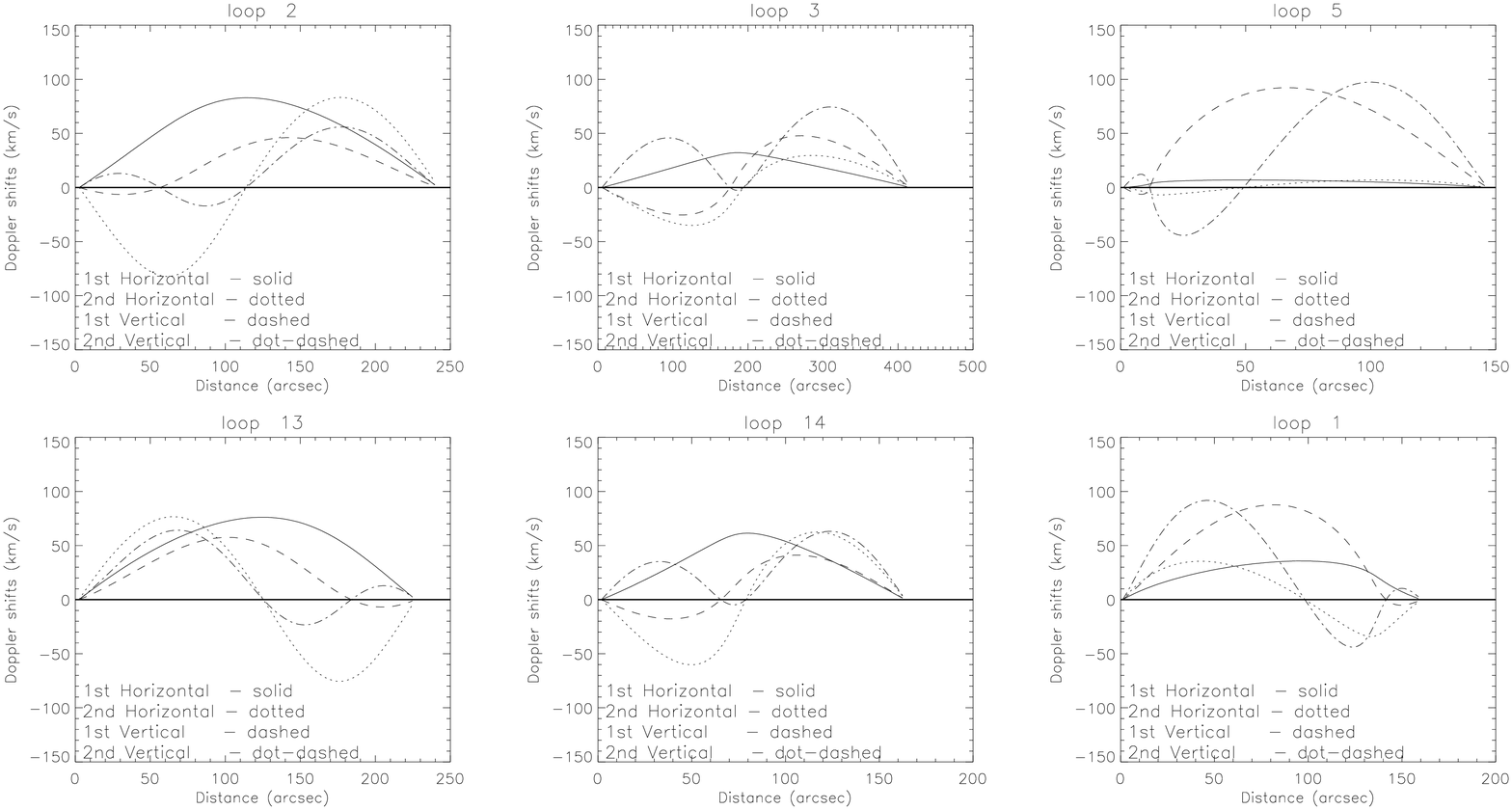}
\caption{Comparison of the fundamental horizontal (solid line), the second-harmonic horizontal (dotted line),
the fundamental vertical (dashed), and the second-harmonic vertical (dot-dashed) oscillations in the 
line-of-sight Doppler shift for three categories of loops: I ({\em left panels}, see Fig.~\ref{type1}), 
II ({\em middle panels}, see Fig.~\ref{type2}) and III ({\em right panels}, see Fig.~\ref{type3}). 
The maximum amplitudes of the absolute value of velocity along the projected loop for four types of 
kink modes are normalized to be 100 km~s$^{-1}$ in order to facilitate the comparison of Doppler 
shift profiles. Positive Doppler velocity represents 
a redshift. On the horizontal axis 0 corresponds to the left footpoint of the oscillating loop.}
\label{doppler}
\end{figure*}

\noindent
{\em -- Category II (Loops 3, 6, 8, 9, 11 and 14):}\\
Most of the loops in this category are located at the limb, except for loop 14, which is located relatively near
the limb. The two typical examples shown in Fig.~\ref{type2} reveal that the fundamental horizontal 
and the second-harmonic vertical oscillations look similar. Both exhibit a cross-over between  
maxima of the displacement near the loop top. The fundamental vertical oscillation 
is distinctly different, exhibiting a lateral displacement with the maximum amplitude close 
to the loop top. Based on these signatures, 
\citet{wan04b} identified the fundamental vertical kink oscillations of a limb loop in the TRACE 195~\AA~band.
The second-harmonic horizontal oscillation may also be easily distinguished from other types of oscillations.
It shows evident displacements at both legs,  so that the loop alternates between thin and fat shapes.  
\citet{asc02} showed that the oscillations in all 6 loops in this category are most evident in the loop legs,
which is consistent with the signature of the horizontal oscillation or the second-harmonic vertical oscillation. 
However, the difference images shown in their paper did not clearly reveal cross-over as expected for the
fundamental horizontal or the second harmonic vertical mode. Neither did it clearly indicate the change in loop
shape expected for the second harmonic horizontal mode. This may be because the 
two images they used were not taken at the times when the loop is located at the maximum and minimum 
positions of the oscillations, respectively. The other reason could lie in the fact that in most
cases the oscillatory motion of the loop is superimposed on a trend motion (a systematic motion) of
the entire loop. Therefore, the expected signatures may be difficult to see. 
However, the second-harmonic horizontal oscillation should be distinguishable from the other
two similar modes, i.e. the fundamental horizontal and the second-harmonic vertical modes, by analyzing
the phase relationship of the oscillations at the two loop legs. For the former the displacements are
in anti-phase while for the latter the displacements are in phase. 

Due to the reasons mentioned above, we re-examined the TRACE data for the 6 loops in this category. 
The difference image plotted in Fig.~\ref{loptrc}a shows that loop 9 has an initial displacement most 
prominent at the loop top, which is consistent with the vertical motion, while the displacements in the 
following period seem to agree with the combination of the fundamental horizontal and 
vertical oscillations (see Sect.~\ref{shvsct}). For loops 3, 8 and 14  the second-harmonic horizontal 
oscillation can be excluded because their two legs indicate the in-phase motions. For loops 6 and 11
only one of their legs is visible, so that it is only possible to rule out the fundamental vertical mode. 
The identified possible modes for the 6 loops are listed in Table~\ref{partab}. We shall show 
in Sect.~\ref{dplsct} that the Doppler shift observations may help distinguish the fundamental horizontal 
from the second-harmonic vertical modes, which exhibit very similar spatial signatures.

\begin{figure}
\centering
\includegraphics[width=8cm]{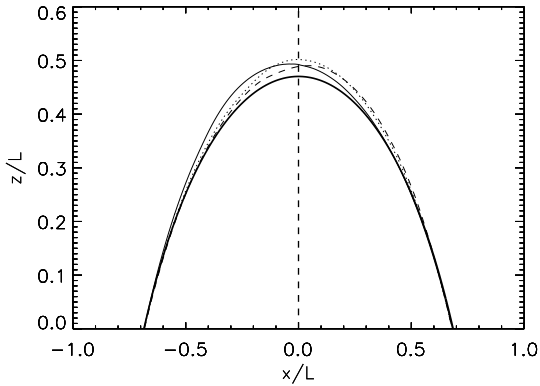}
\caption{MHD simulations of swaying-like loop oscillations excited by a hot pulse launched at
($x_0/L$,$z_0/L$)=($-$0.42,0) where $L=100$ Mm (for details see Selwa et al. 2006).
The curves indicate the positions of
a field line in the oscillating loop at a series of times: t=0 (thick solid line),
t=200 s (thin solid line), t=300 s (dotted line), and t=375 s (dashed line).}
\label{simdist}
\end{figure}

\noindent
{\em -- Category III (Loops 1 and 5):}\\
Those loops are located on the solar disk with the loop plane inclined only slightly relative to 
the line of sight. Figure~\ref{type3} shows two such cases. In this category the fundamental and
second-harmonic vertical oscillations show a similar signature in the sense that the oscillation 
is evident only in a small part, near one end, of the projected loop. It is easily possible to miss
such a weak oscillation signature in observations. The horizontal oscillation 
instead shows clear horizontal displacements over a large part of the loop. Its second harmonic
can be easily identified based on a node at the loop apex, for which the displacements on both sides
oscillate in anti-phase. Based on the observed signature shown by \citet{asc02} we identify the 
transverse oscillations of loops 1 and 5 as the fundamental horizontal kink mode.

\subsection{Doppler features}
\label{dplsct}
Spectroscopic observations provide additional information (e.g., Doppler shifts) that may help distinguish 
between various types of kink modes. Observations of the kink-mode loop oscillations with TRACE thin filters 
have indicated that the loops did not experience significant heating or cooling during the oscillations, 
because they were usually seen to exist before and after the events \citep{sch02, asc02}. Thus, it should be
possible to select certain coronal lines with a suitable formation temperature to detect Doppler shift signatures
of kink modes in coronal loops and hopefully to distinguish between the different possible modes of oscillation. 
Here we calculate for four types of kink modes the distribution of Doppler velocity along the loop for 
a snapshot corresponding to the time of the loop's passage through its equilibrium position. 
At that time the perturbed velocity of the loop reaches its maximum. The calculated LOS Doppler velocity
distributions for an example taken from each of the three categories of loops are shown in Fig.~\ref{doppler}.
Positive velocity represents redshifts. For comparison purposes, the maximum amplitude of the absolute 
value of velocities along the loop has been normalized to 100 km~s$^{-1}$ for all modes of oscillation.

\begin{figure*}
\centering
\includegraphics[width=16cm]{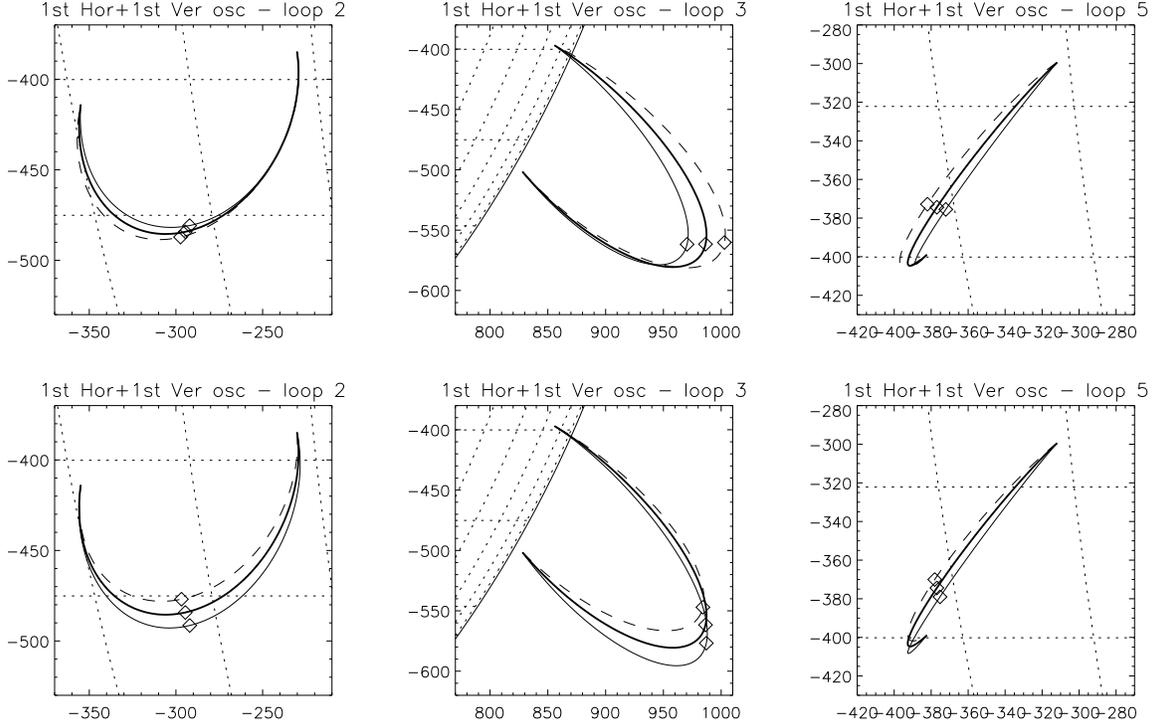}
\caption{Spatial signatures of the combined fundamental horizontal and vertical modes for three loops, 
each observed from a different direction. For the two polarization components the same maximum relative 
amplitudes are assumed and two
phase relations are considered. ({\em Top panels}) Case A - The dashed line represents the extreme position when the loop 
is expanding while horizontally moving northward and the thin solid line the extreme at opposite phase. 
({\em Bottom panels}) Case B - The horizontal oscillation component keeps the same phase but the vertical oscillation 
component has an opposite phase to the above case.}
\label{lophv}
\end{figure*}

We first consider a loop belonging to Category I (e.g., loops 2 and 13 whose Doppler shift is plotted in 
left panels of Fig.~\ref{doppler}). 
The fundamental horizontal oscillation shows a redshift over the entire loop, with the peak redshift near the 
middle of the projected loop. The fundamental vertical oscillation shows a redshift over the major part 
of the loop and a weak blueshift near one end. The second-harmonic horizontal and vertical oscillations
both show sinusoidal redshifts over half the loop, but display quite different profiles of Doppler shifts over
the other half. Therefore Doppler shift measurements can help distinguish different types of kink 
modes, especially among those types of oscillations which show a similar spatial signature in this 
category (fundamental and second harmonic horizontal modes). For that purpose it is sufficient to
measure the Doppler shift at two positions in the loop, one in each half.

In Category II (e.g., loops 3 and 14 in Fig.~\ref{doppler}, {\em middle panels}), the fundamental horizontal
oscillation shows a similar Doppler signature as in Category I but the peak appears slightly
sharper. The Doppler shift signatures of the second-harmonic horizontal and the fundamental vertical 
oscillations look similar to each other, but different to those of the other modes. The second-harmonic 
vertical oscillation is characterized by two completely separated (due to a node at the apex) redshift peaks. 
The fundamental horizontal and the second-harmonic vertical oscillations, which are difficult to 
distinguish based on the spatial signature (see Sect.~\ref{spcsct}), can be well distinguished based on 
the Doppler signature. Again, measurements at two locations in the loop, one in the leg, one at the apex,
should be sufficient.
 
Finally, in loops belonging to Category III (loops 1 and 5 in Fig.~\ref{doppler}, {\em right panels}),
the horizontal oscillations in both the fundamental and second-harmonic modes have very small amplitudes
in Doppler shift due to their motions almost perpendicular to the line-of-sight. Therefore, their
Doppler shift oscillations are difficult to be detected. In contrast, the
vertical oscillations display large amplitudes in Doppler shift and their fundamental and second-harmonic
modes can be easily distinguished based on the Doppler shift profiles along the loop. 
As we have seen in Sect.~\ref{spcsct} exactly these two modes are difficult to detect in imaging observations.
Therefore, the imaging and spectroscopic detections of the horizontal and vertical oscillations
are complementary for this category of loops. 

\begin{figure*}
\centering
\includegraphics[width=16cm]{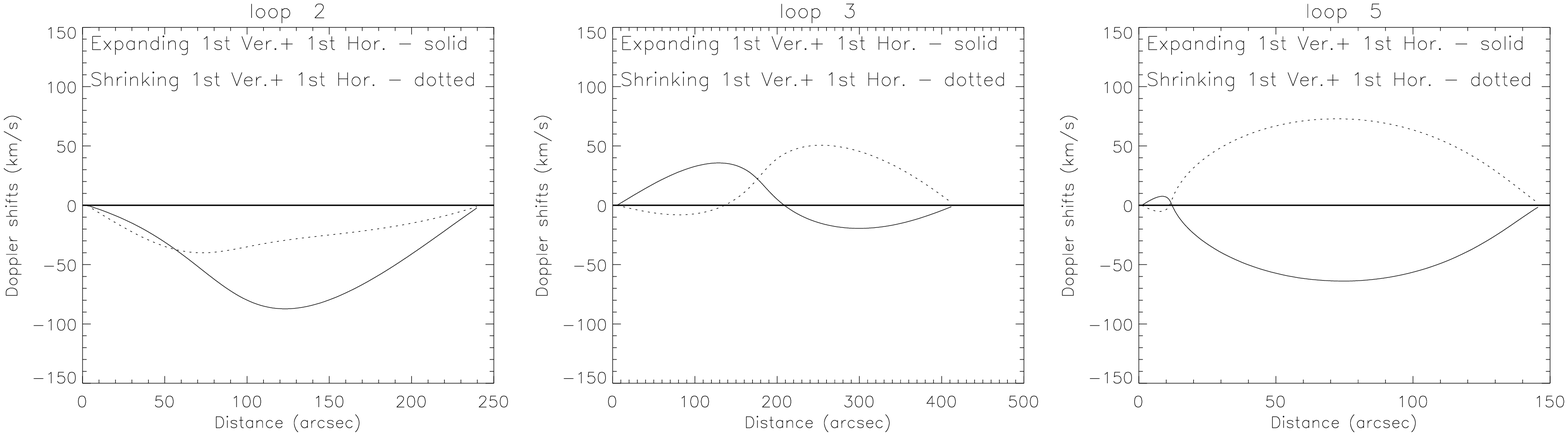}
\caption{Doppler shift signatures of the combined fundamental horizontal and vertical modes for the loops 
plotted in Fig.~\ref{lophv}. The solid curve represents the case shown in case A (the top panels of 
Fig.~\ref{lophv}), while the dotted curve for case B (the bottom panels of Fig.~\ref{lophv}).}
\label{velhv}
\end{figure*}

\section{Simultaneous presence of multiple (harmonics) modes}
\label{shvsct} 

Another reason, besides an inappropriate direction of view, for the failure to identify 
the mode of oscillation is the possibility of multiple modes being excited 
in the loop. Such a case would result in more complex features and would make the distinction based on our idea 
more complicated or sometimes impossible when considering a combination of the fundamental and the second harmonic
modes of the same polarization (e.g., for loops 4, 6, 7, 10 and 11) or when considering a combination of the 
fundamental horizontal and vertical modes (see the following discussion). Two events of the second harmonic 
of the horizontal kink mode were reported by \citet{ver04} in a post-flare loop arcade. However, both of them 
existed together with the fundamental mode. \citet{van07} also found a kink oscillation event showing two wave 
periods, which are consistent with the periods of the fundamental and the second harmonic modes, respectively. 
No pure higher harmonic mode has been observed so far. The simultaneous excitation of the fundamental and higher
harmonics for the kink modes is also suggested in a theoretical analysis by \citet{ter07}.

\begin{figure*}
\centering
\includegraphics[width=16cm]{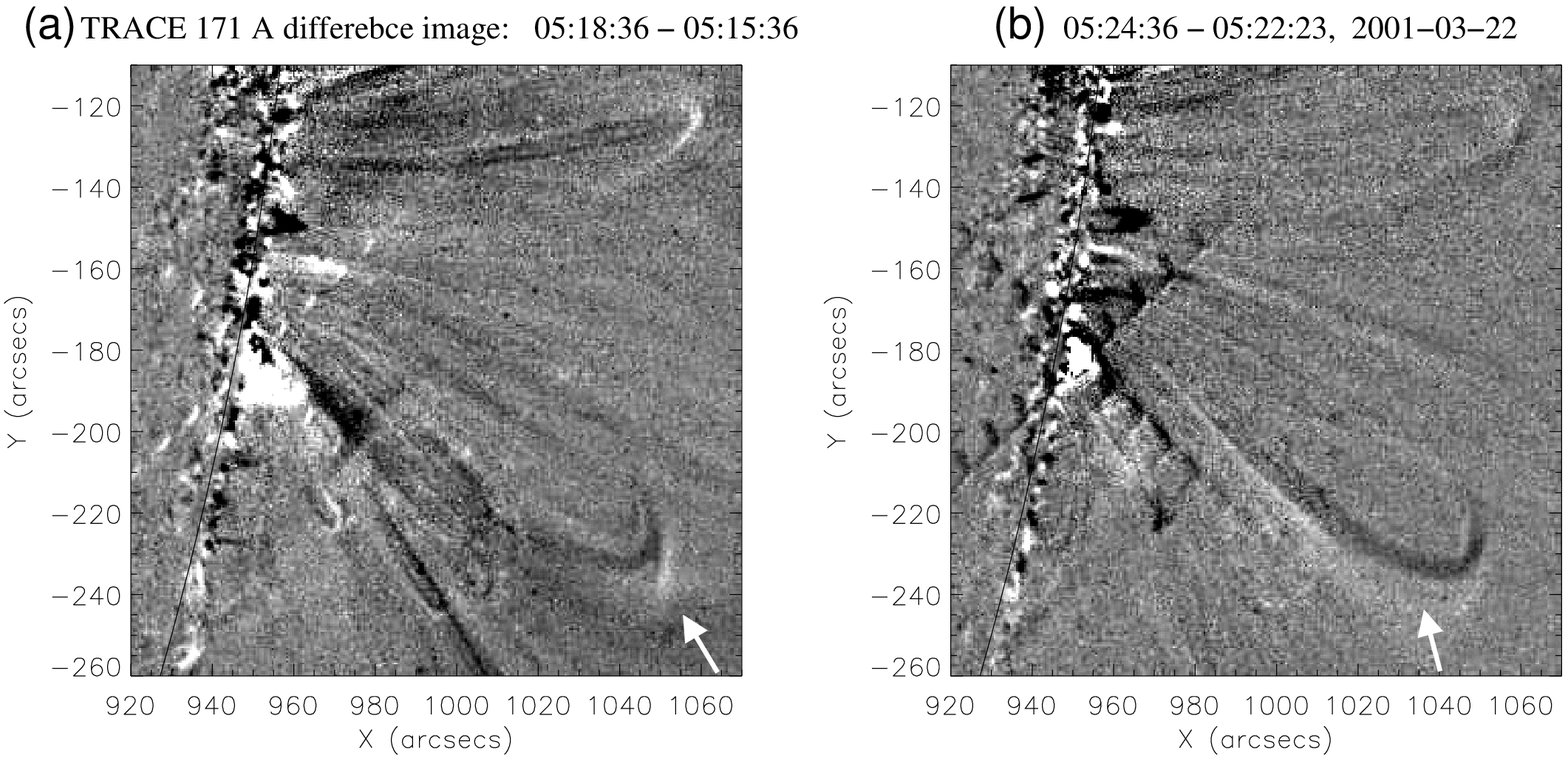}
\caption{ Transverse oscillation of loop 9 seen in TRACE 171 \AA\ difference images. The dark regions
 represent the loop seen in the earlier map and the white regions represent the loop in the current map. 
Evident motions of the loop are indicated by a white arrow. }
\label{loptrc}
\end{figure*}

A good numerical example of simultaneous excitation of both the 
fundamental and the second harmonic mode (in case of vertical kink oscillations) can be provided by 
reconsidering the result of simulations described by \citet{sel06}. 
Based on a 2D MHD arcade loop model, \citet{sel06} showed that a pulse launched near one of the 
footpoints can excite multiple oscillation modes, in particular a swaying-like (possible 
second-harmonic vertical) mode overlaid on a fundamental vertical mode that results in shifts of the apex. 
Figure~\ref{simdist} shows the positions of a field line 
within the oscillating loop at four points in time. We notice a sideways motion of the upper part of the 
loop. The ratio of the oscillation period of the swaying-like mode to that of the fundamental vertical mode, 
which is measured to be about 0.75, agrees well with the theoretically predicted value \citep[see Table 1 in]
[where the second-harmonic vertical oscillation is termed the swaying mode]{dia06}. This indicates 
that the swaying signature seen in the simulation may be caused by the overlaping of the fundamental and 
the second-harmonic vertical modes. Moreover, \citet{sel06} found that the swaying-like mode dampens 
less efficiently than the fundamental vertical mode. This feature is consistent with the property of energy
leakage as the dominant damping mechanism, i.e., the damping is weaker for higher harmonics 
\citep[e.g.,][]{ver06a}.

\begin{table*}
 \begin{center}
\caption{Comparison of spatial and Doppler shift signatures for four types of single kink mode and the combined 
fundamental horizontal and vertical modes$^{\mathrm{a}}$. 
}
\label{cmptab}
\begin{tabular}{lllll}
\hline
Category & Definition & Spatial signatures & Doppler shift signatures & Examples \\
\hline
   & loops located near disk & 1H, 2H and 1V look similar,   & 1H ans 2H can be easily & Loops 2, 4, \\
I  & center and loop plane inclined & while 2V is distinctly  & distinguished, but 1H and  & 7, 10, 12, 13\\
   & to LOS by $~40^{\circ}-60^{\circ}$ & different from others.  & 1V cannot be. & \\
   &                              &  In case A, 1H+1V can be distinguished &  1H+1V and 1V can  be & \\
   &                              &  from the four single modes. In  &  distinguished in case B. & \\
   &                              &  case B, 1H+1V and 1V look similar.         &  &\\
\hline
   & loops located near  &1H and 2V look similar, and & 1H and 2V can be easily & Loops 3, 6, \\
II & the limb            & they may be distinguished from & distinguished. 2H and 1V & 8, 9, 11, 14\\
   &                     & 2H, while 1V is distinctly different.  & look similar. & \\
   &                     &  In both cases 1H+1V can be easily  &   1H+1V may look similar & \\
   &                     &  distinguished from the four       &   to 2H or 1V. & \\
   &                     &  single modes. &  &\\
\hline
    & loops located on disk & 1V and 2V look similar, and are & 1H and 2H are poorly & Loops 1, 5\\
III & and loop plane nearly & hard to detect, but can be easily & detectable. 1V and 2V can & \\
    & parallel to LOS    & distinguished from 1H and 2H.       & be easily distinguished. & \\
    &                    &  In both cases 1H+1V looks   &   1H+1V can be easily &\\
    &                    &  similar to 1H.              &   distinguished from 1H. & \\                
\hline
\end{tabular}
\end{center}
\begin{list}{}{}
\item[$^{\mathrm{a}}$] 1H and 2H indicate the fundamental and the second-harmonic horizontal oscillations, 
respectively, 1V and 2V the fundamental and the second-harmonic vertical oscillations. 
1H+1V indicates a combination of the fundamental horizontal and vertical modes. The definitions for cases A
and B are given in Sect.~\ref{shvsct}.
\end{list}
\end{table*}

Since the fundamental horizontal and vertical modes are the most probable eigenmodes of the kink oscillation
and have very similar periods and damping times \citep{ter06}, it is likely that the two polarizations
are excited simultaneously depending on the exciter and how the blast wave interacts with the loops. 
As a comparison with the single modes, we calculate the spatial and Doppler shift
signatures of the combined fundamental horizontal and vertical modes (see Figs.~\ref{lophv} and~\ref{velhv}).
The same relative amplitudes are assumed for the two polarization components, i.e., $A_m=\phi_m=0.062$ (arc deg),
so that the total maximum relative amplitude is about $\sqrt{2}A_m=0.087$, same as used to model the
single modes in Sect.~\ref{trcsct}. Two cases are considered, in which the vertical oscillation
components have the opposite phase while the horizontal oscillation component retains its phase. 
We refer to the case shown in the upper panels of Fig.~\ref{lophv}
as case A, and the case in the bottom panels as case B. The main spatial and Doppler shift characteristics 
are summarized in Table~\ref{cmptab}.

The loops belonging to Catogory I show an evident displacement at one leg 
in the projected plane in case A (e.g., Fig.~\ref{lophv} {\em top left}), which can be easily 
distinguished from the four types of kink modes analyzed in Sect.~\ref{spcsct}. Note that the fundamental 
and second harmonic horizontal modes display an evident displacement at the other leg. However, 
in case B they exhibit an evident displacement around the loop top (e.g., Fig.~\ref{lophv} {\em bottom left}), 
which looks very similar to that for the fundamental vertical mode. 
Therefore, we cannot exclude the possibility that the fundamental vertical oscillation which
has been identified for loops 12 and 13 is combined with the fundamental horizontal mode in this case. 
By comparing their Doppler shift signatures (the dotted curve in Fig.~\ref{velhv}~{\em left} and the 
dashed curve in Fig.~\ref{doppler}~{\em top left}), we find that their peak shifts are located at
different sides of the projected loop. Thus the combined fundamental horizontal and vertical modes may
be distinguished from the single fundamental vertical mode based on the measurement of the Doppler shift
at two positions in the loop, one in each half in case B.

For the loops in Category II, we find that the combined fundamental horizontal and vertical modes can be
uniquely identified, because they display an evident displacement off the loop top in both cases (e.g., 
Fig.~\ref{lophv} {\em middle}), which are clearly different from the features of the four types of kink modes 
analyzed in Sect.~\ref{spcsct}. The TRACE difference images reveal that for loop 9 its initial apparent
displacement is consistent with vertical motion 
(Fig.~\ref{loptrc}a), while the displacements at the following phase consistent with the combined fundamental 
and vertical oscillations (comparing Fig.~\ref{loptrc}b with Fig.~\ref{lophv} {\em bottom middle}). 
It is also possible to distinguish between combined modes with different phase relationships on the basis 
of the skew direction of their evident displacements.

For the loops in Category III, both cases of the combined fundamental horizontal and vertical modes 
show the feature of horizontal displacements very similar to the single fundamental horizontal mode,
but they can be easily distinguished from the horizontal mode based on the large Doppler shifts
along the loop (Fig.~\ref{velhv}~{\em right}). 

\section{Discussion and Conclusion}
We have modeled the geometric distortion to a simple circular loop produced by four types of kink modes 
(fundamental and second harmonic modes of horizontal and vertical oscillations) and considered if it 
is possible to distinguish 
between them for a given observational geometry. We analyzed in detail 14 selected TRACE loops observed 
to harbor transverse oscillations and found that they can be divided into three categories based on the 
location of the observer relative to the loop plane (viewing geometry). The apparent similarity and difference 
between the four types of kink modes in spatial signatures is found to depend strongly on the category. 
We also examined if the kink modes displaying a similar spatial signature can be distinguished based on their Doppler 
shift signatures.  Our main conclusions regarding the possibility of distinguishing between modes are
outlined below (see also Table \ref{cmptab}).

(1) For those loops whose loop plane is almost in the line-of-sight direction (i.e. loops in Category III),
the fundamental and second-harmonic horizontal kink modes can be easily and uniquely identified based on the 
signature of evident lateral displacements of the loop. The fundamental and second-harmonic vertical kink modes 
show smaller displacements with similar signature in this viewing geometry and are difficult to detect and 
distinguish. It is also impossible to determine if a fundamental horizontal mode is combined together 
with a fundamental vertical mode in this case. For those loops located on or near the limb 
(e.g. Category II loops), the fundamental vertical kink mode as well as its combination with the fundamental
horizontal mode are most easy to distinguish from the other types of kink modes. The fundamental
horizontal and the second-harmonic vertical modes look similar, displaying a cross-over signature near the
loop top. For those loops on the solar disk whose loop plane is inclined at an angle 
sufficiently different from $0^{\circ}$ and $90^{\circ}$ to the line-of-sight direction (e.g., the loops 
in Category I), the second-harmonic vertical mode is most easy to be uniquely identified based on the 
cross-over signature near the loop top. The fundamental horizontal and vertical oscillations may be 
distinguished from each other by considering more subtle effects, e.g., determining on which side of the 
loop relative to the loop apex the largest amplitude displacement is located and comparing it with 
a simple model of the type considered here.  However, it may be difficult to determine if the identified 
fundamental vertical mode is present together with the fundamental horizontal mode. 

(2) The Doppler shift signatures provide additional constraints that can help 
distinguish between two types of kink modes which exhibit similar spatial oscillation signatures if the
used spectrometers have a high resolution in Doppler velocity and if the spectrometer slit crosses 
the loop at two or more locations (or if the loop is scanned sufficiently rapidly) and 
if the studied loop is well isolated from the ambient loops. 

(3) The re-examination of the swaying-like motions of the loop top observed in a 2D arcade loop simulated
by \citet{sel06} shows several features such as the period ratio and weak damping relative to the 
fundamental mode that are consistent with the theoretically expected ones for the vertical second 
harmonic oscillation. They could be evidence for the swaying-like motions resulting from the 
second-harmonic vertical mode in the loop. However, there are differences that suggest that the
fundamental vertical mode was also excited. Simultaneous excitation of multiple modes complicates the
identification further. The simple analysis presented here may not be adequate in such cases.

In Sect.~\ref{dplsct} we only considered the shape of velocity profile along a loop, but the
absolute amplitude of Doppler velocity may also be used to distinguish between different kink modes. The 
amplitude of Doppler velocity can be inferred from the apparent displacement amplitude divided by the measured 
oscillation period for a given geometry and oscillation mode. The amplitude is expected to be 
a strong function of the oscillation mode and could be used to identify the mode by comparing with 
direct measurements of Doppler velocity obtained with a slit spectrograph at only a single location.

We have determined which type of kink modes each of the 14 transverse loop oscillations observed by TRACE
belongs to based on a comparison between the observed and modeled signatures 
(see the last second column in Table~\ref{partab}). For 2 loops out of the 6 loops in Category I the oscillation 
corresponds to the fundamental vertical mode  (but may be combined with the fundamental horizontal mode), 
and for another it is identified as the fundamental horizontal kink mode. The remaining 3 could not be 
uniquely identified.  Out of the 6 loops in Category II one loop shows the interesting feature which
first indicates the dominant vertical motion and then agrees with the combined fundamental vertical and horizontal modes.   
The oscillations of the other loops could not be uniquely identified. The two loops in Category III are clearly 
identified as displaying the fundamental horizontal kink mode. The results suggest that horizontal
oscillations seem to be more easily excited under solar condition. This may be because the disturbing
sources (e.g., flares) are commonly located near to but rarely immediately in the plane of the
oscillating loops. Note, however, that any vertical oscillations in loops of Category III would not easily 
have been detected in TRACE data.

In summary, our analysis has shown that for a given viewing geometry it may be difficult to distinguish
between some of the different kink modes and in future when identifying kink modes the results of this work should 
be taken into account, and comparisons should be made to simple geometrical models, like those considered here. 
Ideally, the application of this method requires the complete loop to be well visible. This is not always 
the case with observed oscillating loops, which are sometimes only partially visible.

The current analysis is not complete. Thus, it is restricted to loops lying in a plane and does not consider
the case of sheared loops such as that reported by \citet{dem07}. We have also not taken into account 
the information in the oscillation period, although it is affected by different atmospheric parameters
and may not always help distinguish between modes. Another way of distinguishing between vertical and 
horizontal modes may be through brightness fluctuations \citep{wan04b}, but before doing that improved 
statistics of vertical oscillations are needed. Finally, stereoscopy, when corresponding data are
available, should be very helpful for obtaining the true geometry of oscillating loops and of the oscillations
they harbor \citep[e.g.][]{fen07}.

\acknowledgements
TJW's work was supported by NRL grant N00173-06-1-G033, NASA grant NNG06GA37G, and NASA grant NAS5-38099 
for TRACE mission operations and data analysis, through a subcontract of Lockheed-Martin Solar and Astrophysics 
Laboratory with Montana State University. MS's work was supported by the NASA grant SEC theory program
and NASA grant NNG06GI55G. The authors also thank the anonymous referee for his constructive comments.

\end{document}